\documentclass[]{spie}  

\usepackage{amsmath} 
\usepackage{alltt}
\usepackage{array}
\usepackage{epsfig}
\usepackage{verbatim}
\usepackage{fancybox}
\usepackage{multirow}
\usepackage{colortbl}
\usepackage{color}
\usepackage[table]{xcolor}
\usepackage{algorithm}
\usepackage{algorithmic}
\usepackage{url}
\usepackage{soul}
\usepackage{hyperref}
\usepackage{booktabs}
\usepackage{subcaption}
\title{Can Multimodal Sensing Detect and Localize Transient Events?}
\author[]{Kasthuri Jayarajah$^{\ddagger}$}
\author[]{Vigneshwaran Subbaraju$^{\ast}$}
\author[]{Noel Athaide$^{\dag}$}
\author[]{Lakmal Meegahapola$^{\ddagger}$}
\author[]{Andrew Tan$^{\ddagger}$}
\author[]{Archan Misra$^{\ddagger}$}
\affil[]{$^{\ddagger}$Singapore Management University, Singapore}
\affil[]{$^{\ast}$Agency for Science Technology and Research (A*STAR), Singapore}

\authorinfo{$^{\dag}$ Work carried out while a Senior Research Engineer at Singapore Management University. Further author information: (Send correspondence to A. M.)\\A.M.: E-mail: archanm@smu.edu.sg, Telephone: $+$65 68085202\\
K.J.: kasthurij.2014@smu.edu.sg, V.S.: vigneshwaran\_subbaraju@sbic.a-star.edu.sg,\\ N.A.:athaiden@gmail.com, L.M.: lakmalm@smu.edu.sg, A.T.: andrewtan@smu.edu.sg\\
}

\begin{document} 
\maketitle

\begin{abstract}
With the increased focus on making cities ``smarter", we see an upsurge in investment in sensing technologies embedded in the urban infrastructure. The deployment of GPS sensors aboard taxis and buses, smartcards replacing paper tickets, and other similar initiatives have led to an abundance of data on human mobility, generated at scale and available real-time. Further still, users of social media platforms such as Twitter and LBSNs continue to voluntarily share multimedia content revealing in-situ information on their respective localities. The availability of such longitudinal multimodal data not only allows for both the characterization of the dynamics of the city, but also, in detecting anomalies, resulting from events (e.g., concerts) that disrupt such dynamics, transiently. In this work, we investigate the capabilities of such urban sensor modalities, both physical and social, in detecting a variety of local events of varying intensities (e.g., concerts) using statistical outlier detection techniques. We look at loading levels on arriving bus stops, telecommunication records and taxi trips, accrued via the public APIs made available through the local transport authorities from Singapore and New York City, and Twitter/Foursquare check-ins collected during the same period, and evaluate against a set of events assimilated from multiple event websites. In particular, we report on our early findings on (1) the spatial impact evident via each modality (i.e., how far from the event venue is the anomaly still present), and (2) the utility in combining decisions from the collection of sensors using rudimentary fusion techniques.
\end{abstract}
\section{Introduction}
An extraordinary amount of urban transportation and mobility data are being increasingly collected, and made publicly available, by cities across the globe. Such data encompasses both public/municipal agencies and private companies (e.g., telcos and social media). Examples include (a) public transportation-related data such as the trajectories and occupancy levels of buses, the GPS-derived movement trajectories \& trip records of taxis and the occupancy levels of parking garages; (b) large-scale, individual-level movement patterns, obtained from Call Detail Records, by telecom companies; and (c) user-generated text, image and location data (e.g., via check-ins) posted on social media platforms such as Foursquare and Twitter. Such data effectively provide multiple observation modes to capture the collective mobility dynamics of a city.


Although an abundance of multi-modal data is now available to study urban mobility, there's a lack of systematic studies that investigate the utility in relying on such multiple modes. For instance, many past works that study urban dynamics relied on single sources -- e.g., taxi-data based studies \cite{chawla2012inferring, liu2011discovering}, social media-data based studies \cite{Noulas:2015, GeorgievNM14b} -- and tend to treat mobility data sources as \emph{substitutes}, but the fundamental question of whether they are in fact perfect substitutes, or is there utility in \emph{combining} multiple sensor modes, is largely unexplored. In this work, we take the case of urban event detection, a key problem extensively researched in recent years \cite{jayarajah2017discovering, giridhar2016clarisense+, chawla2012inferring, liu2011discovering}. Our conjecture is that different individual modalities differ in their discriminatory power, and that they exhibit different levels of spatiotemporal sensitivity to urban disruptions. For instance, it is natural to assume that most people will arrive at an event venue prior to its start time, and that the resulting spikes in arrival rates can thus help determine the \emph{onset} of the event. However, it is possible that different transportation modes exhibit slightly different temporal properties--e.g., people using public buses and trains arrive may slightly earlier (even more in advance of the start time) than people using taxis and private vehicles.

\begin{figure}[b]
\centering
\includegraphics[width=5in]{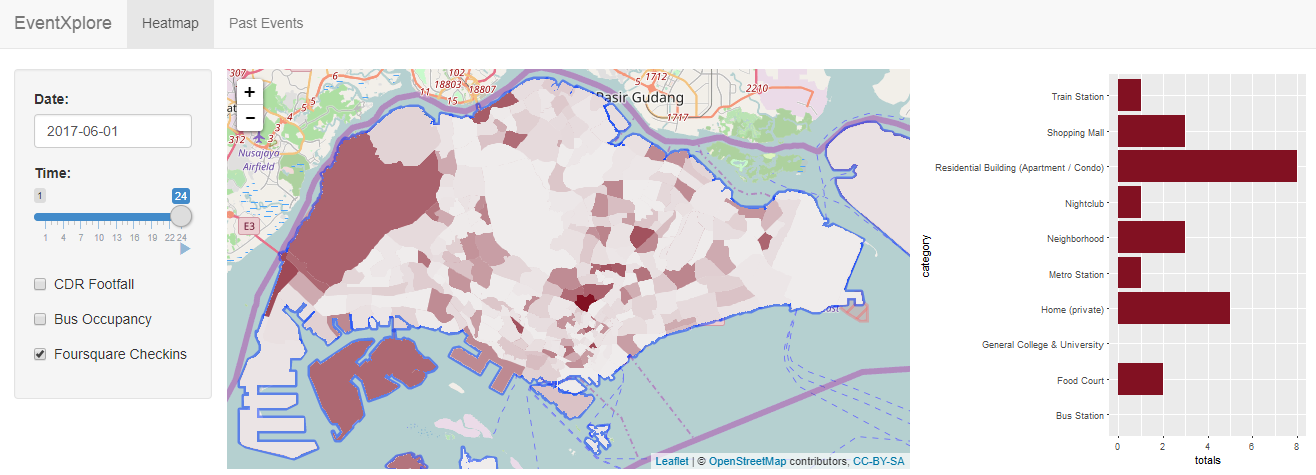}
\caption{Landing Page View of the \emph{EventXplore}.}
\label{fig:dashboard} 
\end{figure}

To this end, we curate multiple transport-related and social media datasets from two cities, Singapore and New York City, and collect a set of over 150 events from a number of news and event websites to answer three key questions: (1) are individual mobility sources (both social and physical) reasonable choices for detecting urban events?, (2) are there differences in the spatiotemporal sensitivity of the individual sources?, and finally, (3) can the use of simple fusion techniques help in improved performance for detection over single-source modalities? In addition to providing insights on these questions, we propose and build EventXplore (illustrated in Figure~\ref{fig:dashboard}), a prototype web application that demonstrates the use of such disparate urban mobility data sources in detecting and visualizing events and semantically annotating them using user-generated textual content on social media. 

In this work, we make the following contributions:
\begin{enumerate}
    \item{Using real world, city-scale datasets from two major cities (Singapore and New York City), we investigate the ability of multiple urban data sources to detect urban events, of varying categories and scales, using statistical outlier detection tests. In particular, using $150+$ events curated across both NYC and Singapore, we demonstrate that data sources differ in their detection accuracy, and in their the ability to localize an event spatially and temporally. We observe recall rates of 30 - 50 \% in locating an event within 1.5 $km$ of its venue, and up to 80\% within 4 $km$. Further, we also find that the sensor modalities are equally good at detecting events \emph{prior} to their scheduled start times, a possible result from the natural tendency of people in \emph{arriving} earlier to an event.}
    \item{Using a number of rudimentary fusion techniques, we explore the utility in fusing individual sensor modalities and conclude that fusing at the decision level (e.g., majority voting over decisions made independently per sensor) performs better than fusing at the data-level (e.g., averaging of outlier scores across the multiple sensors) in the settings we consider. However, the significantly weaker performance of the combination sensors in comparison to the best performing individual sensor highlights the need for better fusion algorithms. }
    \item{We propose a framework for such ingestion and fusion of multimodal urban data and prototype \emph{EventXplore}, a system for detecting events from multimodal data that provides several analytics views and semantic annotation.}
\end{enumerate}

In the remainder of this paper, we first describe the datasets considered in this work in Section~\ref{sec:data}, discuss the pipeline of event detection that we propose in Section~\ref{sec:overview}. We share our current findings in Section~\ref{sec:eval}, discuss limitations and future work in Section~\ref{sec:discussion}, briefly summarize related work and conclude in Section~\ref{sec:related}.
\section{Preliminaries}
\label{sec:data}
In this section, we first describe the datasets considered in this work, and summarize briefly results from statistical tests that demonstrate that the multiple datasets are not substitutes of each other and may have additional value of information than relying solely on single sources of data.

\subsection{Dataset Description}
In this work, we rely primarily on two classes of datasets: (1) physically-sensed and (2) from social media.

\textbf{Physically-sensed data: }In the case of Singapore, we focus on events during the period of May - June 2017. We collected CDR records of a major telecommunications service provider through their APIs at DataSpark \footnote{\url{https://datasparkanalytics.com}}; in particular, we collected information on \emph{Discrete Visits} which is the number of people entering a zone at hourly intervals throughout the day. Each record is a tuple containing $<z, t, u>$ where $z$ is an administrative boundary (i.e., a subzone \footnote{\url{https://data.gov.sg/dataset/master-plan-2014-subzone-boundary-web}}), $t$ is an hourly window and $u$ is the total number of unique users observed at that zone for that period. In addition, we collected occupancy levels of incoming buses for over 260 bus services from over 130 bus stops island-wide using the publicly available DataMall API \footnote{\url{https://www.mytransport.sg/content/mytransport/home/dataMall.html}}. Each record is a tuple $<busstopid, serviceid, timestamp, loading>$ where $busstopid$ and $serviceid$ identify the bus stop and the service, respectively, and $loading$ is a discrete number varying between 1 and 3 capturing the level of crowdedness of the next bus scheduled to arrive at that stop. The API is refreshed every 2-3 minutes, with $timestamp$ capturing the most recent refresh instant.

In the case of New York City, we focused on events that happened during the year of 2013, and we obtain time-stamped records of dropoffs and pickups by yellow taxis for the period of January 2013 - December 2013, made available publicly by the New York City Taxi and Limousine Commission \footnote{\url{http://www.nyc.gov/html/tlc/html/about/trip\_record\_data.shtml}}. Each record is a tuple $<p_x, p_y, t_p, d_x, d_y, t_d>$ where $(p_x, p_y)$ and $(d_x, d_y)$ are the GPS coordinates of the pickup and dropoff of a single taxi trip, respectively, and $t_p$ and $t_d$ are the corresponding timestamps. We aggregate the pickup and dropoff points to the corresponding Census tracts \footnote{\url{http://maps.nyc.gov/census/}},  where a tract typically houses at most 15000-16000 residents \footnote{\url{https://data.cityofnewyork.us/City-Government/2010-NYC-Population-by-Census-Tracts/si4q-zuzm}}.

\textbf{Social Media data: } For both Singapore and NYC, we collect check-ins from the popular location-based social network Foursquare\footnote{\url{https://developer.foursquare.com/}}. For Singapore, we crawl public check-ins that were re-shared as Tweets \footnote{\url{https://developer.twitter.com}} by users identified as being from Singapore, and that were visible publicly. The public Tweets were collected using the Streaming API. In the case of NYC, we use publicly available data \cite{yang2015nationtelescope, yang2015participatory}.
Each check-in is a tuple $<venueid, timestamp, v_x, v_y, v_c>$ where a venue or location is identified by its $venueid$ whose latitude, longitude, and category are $v_x$, $v_y$ and $v_c$, respectively, and $timestamp$ is when a user checked-into that venue. The categories fall within Foursquare's API of hierarchical categories, and can be obtained by querying the Foursquare API \footnote{\url{https://developer.foursquare.com/categorytree}}. Due to the sparsity of venue-level check-ins, we aggregate check-ins to their respective administrative boundary (e.g., subzone or Census tract) in our analyses.

\textbf{Events data:} We focus on a set of 15 manually curated events from various event and news portals from the period of May-June 2017 in Singapore, and over 160 events in New York City during the whole year of 2013. We further filter the list of events to retain only those ``localized" events that were both geographically and temporally confined---for example, public holidays were removed as they typically cause multiple events distributed throughout the day. For multi-day events, we focus on the detection of the start time on the first day of the event. We restrict our analyses to 85 subzones in the greater Central Business District (CBD) in Singapore and 288 Census Tracts within the Manhattan Borough in NYC.

In Table~\ref{tab:data}, we summarize the datasets used in this work.

\begin{table}[!th]
\begin{center}
\resizebox{\textwidth}{!}{
\begin{tabular}{p{2.5cm}  >{\centering\arraybackslash}m{3cm}  >{\centering\arraybackslash}m{3cm} >{\centering\arraybackslash}m{3cm} >{\centering\arraybackslash}m{3cm} >{\centering\arraybackslash}m{3cm}  }
 \hline
\textbf{City} & \textbf{Observation Period} & \textbf{Data Source}  & \textbf{Spatial Resolution} & \textbf{Temporal Resolution} & \textbf{Total Records} \\ \hline
Singapore & May - June 2017 & CDR & subzone & hourly &  124,440\\
 & & Bus Arrival & 260 Bus Stops & 2-3 mins & 16 million \\
 & & Check-ins & 20,801 Venues & continuous & 201,863 \\ \hline
New York City & January - September 2013 & Taxi & continuous & continuous & 143 million trips \\
& & Check-ins & 29,487 venues & continuous & 325,161 \\ \hline
\hline
\end{tabular}
}
\end{center}
\caption{Summary of Datasets Used in this Work.}
\label{tab:data}
\end{table}



\subsection{Why use Multiple Sources?}
To understand whether the use of multiple data sources can in fact provide an information gain, as opposed to relying on a single source, we performed a series of Granger-causality tests \cite{granger1969investigating} between each pair of data source. This test quantifies the ability of a time series $x$ to forecast another time series $y$. Operationally, we compare the power of $x$ in predicting $y$ compared to predicting based on past values of $y$ alone. The $p-$ values from a Wald test between the two models is tabulated in Table~\ref{tab:granger} for the datasets from NYC as an illustration. A $p-$value $\geq$ 0.05 means we reject the null hypothesis that the time series $x$ has statistically significant explanatory power over time series $y$. We see that in all cases that this is in fact the case, and that the multiple data sources may capture different aspects of urban mobility. This is an important finding, as it indicates the likely insufficiency of prior methods that have focused on detecting events solely using one type of data source.

\begin{table}[!th]
\begin{center}
\resizebox{0.5\textwidth}{!}{
\begin{tabular}{p{2cm}  >{\centering\arraybackslash}m{2cm}  >{\centering\arraybackslash}m{2cm} >{\centering\arraybackslash}m{2cm} }
 \hline
\textbf{} & \textbf{Checkins} & \textbf{Dropoff}  & \textbf{Pickups}  \\ \hline
\textbf{Checkins} & -& 0.605 (0.292)& 0.599 (0.287)\\
\textbf{Dropoff} & 0.565 (0.301)& -& 0.585 (0.291)\\
\textbf{Pickup} & 0.611 (0.286)& 0.581 (0.299)& -\\
 \hline
\hline
\end{tabular}
}
\end{center}
\caption{Summary of p-values from Granger Causality Tests. The average (std. dev) of p-values, all $\geq 0.05$,  show that the null hypothesis that the individual data sources can forecast each other can be rejected at 5\% confidence level.}
\label{tab:granger}
\end{table}

\section{Framework Overview}
\label{sec:overview}
\begin{figure}[t]
  
  \begin{subfigure}[b]{0.6\textwidth}
  \centering
    \includegraphics[width=4in]{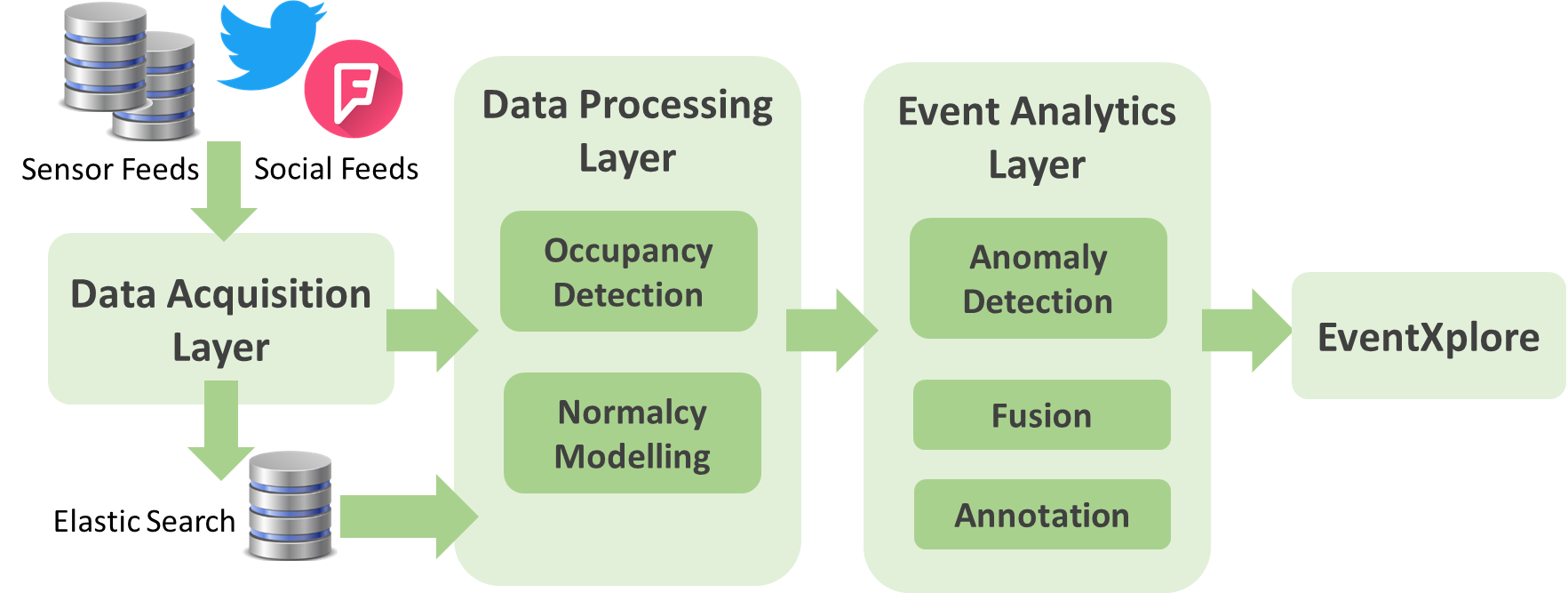}
    \caption{System Overview.}
    \label{fig:sysoverview}
  \end{subfigure}%
  \begin{subfigure}[b]{0.4\textwidth}
  \centering
    \includegraphics[width=2in]{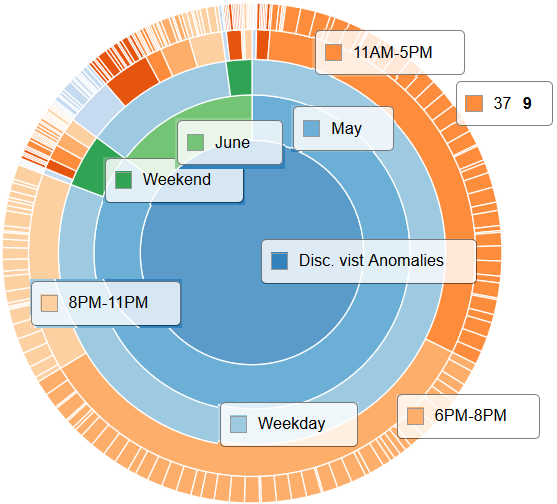}
    \caption{Sunburst view of anomalies}
    \label{fig:sunburst}
  \end{subfigure}
\end{figure}

In the design and implementation of our framework, we focus on the following key components illustrated in Figure~\ref{fig:sysoverview}.

\textbf{Data Acquisition Layer: }This acts as the  unifying layer responsible for the curation of data (e.g., via API requests from DataSpark and DataMall, crawler scripts for Tweets and Foursquare checkins, etc.) whilst handling differences in the various attributes, stipulated query limits and temporal granularity. We use ElasticSearch\cite{Elastic} as the primary data storage service for data sources with large volumes and frequent updates (e.g., bus arrival data is refreshed every 2-3 minutes).

\textbf{Data Processing Layer: } At this stage, two key functionalities are performed. The \emph{Occupancy Detection} component, extracts the \emph{occupancy level} for each data source, the definition of which varies from source to source. In our setting, the occupancy level of CDR, bus arrival, taxi and Foursquare were extracted as the number of discrete visitors, average loading level across all services served by a bus stop, number of taxi pickups/dropoffs, and the number of unique users checking in, respectively. For each source, the aggregated occupancy level $c_{s,w,d}$ at location $l$, during window $w$, on day type $d$ was summarized. For example, for CDR data, the set of all $s$ were the 85 subzones whereas for bus data, this was a sample of 130 services spanning 260 bus stops within the 85 zones. The window $w$ was hourly in the case of CDR (limited by the granularity at which the API provides data), whereas we considered bins of 15 minutes for the remaining sources. The day type was used to differentiate between weekdays and weekends. We assume that $c_{s,w, d}$ follows a normal distribution similar to past work \cite{barria2011detection} and confirm this using the Shapiro-Wilk test for normality with data corresponding to each Census tract and hour of the day (across all weekdays) over the entire period of observation considered to be samples from the same distribution. For instance, in the NYC case, 95.63\% and 83.15\% of the resulting $p-$values were greater than 0.05 suggesting that the null hypothesis that the distributions are normal cannot be rejected at 5\% significance level for taxi dropoffs and pickups, respectively. However, similar evidence was not found for check-ins and we acknowledge the need for alternative outlier representations. We defer such investigations to future work. The \emph{Normalcy Modeling} component computes statistical measures such as the mean and standard deviation of $c_{s,w, d}$ and the normalized z-score values for each observation.


\textbf{Event Analytics Layer: }The \emph{Anomaly Detection} component uses statistical outlier detection techniques to declare an observation (i.e., a tuple $<s,w,d>$ from a data source as \emph{anomalous}. We compare two approaches here: (1) based on a static threshold on the z-score, and (2) based on the Seasonal Hybrid ESD (S-H-ESD) \cite{vallis2014novel} on the computed z-scores. The latter has been shown to be good at detecting both global and local anomalies as the algorithm takes into account seasonal variations and trends. We use the implementation available at \url{https://github.com/twitter/AnomalyDetection} in this work. The \emph{Sensor Fusion} component ingests observations tagged as anomalous from the individual data sources to declare a fused decision. We consider a number of baseline fusion methods such as majority voting and averaging. 

The \emph{Semantic Annotation} component mines the categories of Foursquare venues and keywords from Tweets to add semantics to detected events similar to \cite{giridhar2016clarisense+}. For time bins and locations that the system declares as outliers, we mine the types of Foursquare venues that received the most number of check-ins, and hashtags from Twitter that had the highest $TF-IDF$ scores for annotating the anomalies. We defer the evaluation of this component to future work.

\subsection{Web Application \emph{EventXplore}}
The web application was prototyped using RShiny baed on the framework described and entails the following views.

\textbf{Event Landscape View:} This view, shown in Fig \ref{fig:dashboard}, provides the user a quick snapshot of the events landscape all over Singapore in the form of a series of images in time-lapse. The Foursquare venue types that receive the highest number of check-ins during that time are also visualized.

\textbf{Summary of anomalies:} This interactive view, modeled as a Sunburst diagram, is shown in Fig \ref{fig:sunburst}, and it provides a summary of the anomalies observed in the CDR data. The first three concentric rings represents time units such as month, weekday/weekend and time-bins. The last ring represents the subzone and the count of anomalies. The size of a slice in a ring is proportional to the corresponding number of anomalies.

\textbf{Data Layer View:} This view provides the user with an interface to add/remove data sources used to perform the analysis.

\begin{figure}
\centering
\includegraphics[width=4in]{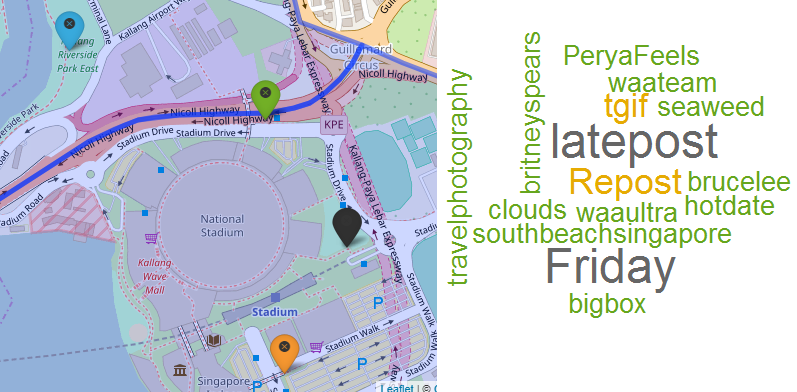}
\caption{\textit{Left} -Estimated and actual event venues of the Britney Spears Concert. Black - Actual, Orange - CDR, Green - Bus, Blue - Foursquare. \textit{Right}-Word cloud during the Britney Spears Concert. Hashtags like ``BritneySpears'' were present.}
\label{fig:combined}
\end{figure}

\textbf{Past Events View:} This view, provides the user with overview of events that were detected in past. In Figure~\ref{fig:combined}, we share a screenshot from the EventXplore dashboard which shows the localization accuracy of each source compared to the actual event venue for one of the largest events from our list -- i.e., the Britney Spears concert. 




\section{Evaluation}
\label{sec:eval}
In this section, we describe our evaluation methodology and share our preliminary findings. We first take the example of CDR data from Singapore and explore the use of short vs. wide observation windows with respect to their ability to detect events of different scales. We then explore the ability of the different sensor modalities in detecting and localizing events spatially and temporally. Finally, we investigate the utility in fusing the multiple modalities using baseline fusion techniques.

\subsection{Ground-truth Events}
To validate our approach, we consulted event-related portals on the internet from both Singapore and New York City, in the months of May and June 2017 for the former, and throughout the year of 2013 for the latter. Based on our search, we selected a few large scale, small scale and medium scale events. Table \ref{tab:events} lists the limited set of events we considered in the context of Singapore which illustrates the type of events we considered for detection. In the evaluation in Section~\ref{sec:spatiotemporal}, we focus on a subset of ``localized" events that were confined in location and duration-- for example, removing holidays as they were typically associated with multiple day-long events, distributed throughout the city. For multi-day events, we focus on the detection of the start time on the first day of the event. 

\begin{table}[t]
  \centering
  \resizebox{0.8\columnwidth}{!}{
  \begin{tabular}{ccll}
    \toprule
    Date&Time&Name&Scale \\
    \hline
    \midrule
    10 May & All day&Vesak Day 3 step 1 bow procession& Large\\
    3-4 Jun & PM peak& Dragon Boat Festival & Medium\\
    10 Jun & All day &  Ultra Singapore Electronic Music Festival  & Medium\\
    17-Jun  & PM peak& Bark and Kisses: A dog cafe adventure & Small\\
    17-18 Jun  & PM peak& Urban Camping  & Small \\
    17 Jun  & All day& Food Expo  & Medium\\
    24-Jun  & All day& Hari Raya Market  & Large \\
    24-Jun & Off-peak& Dreamworks day & Small\\    
    14-Jun & All day& Natl. Inter-School Dragon Boat Championships  & Medium\\
    16-Jun  & PM peak& ADAC 2017 music concert  & Small\\
    30-Jun  & PM peak& Britney Spears music concert  & Large\\
    16-17Jun  & PM peak& OMM:Hensel and Gretel & Medium\\    
    1-4 Jun & All day &  Singapore Intl. Piano Festival   & Small\\       
    9-11 Jun & All day & Health Fiesta & Small\\
    9 Jun  & PM Peak& A-MEI<Utopia 2.0 Carnival> World Tour & Medium\\
    9-11 Jun & All day & Doctors without borders, Sg. Intl. Film Festival   & Small\\
    \bottomrule
  \end{tabular}
  }
\caption{Canonical set of events in Singapore in May/June '17}
\label{tab:events}
\end{table}

\subsection{Choosing the Temporal Granularity}
We first evaluate the ability of CDR data (in Singapore) to localize events with observation windows of several sizes. The analysis was restricted to the 85 subzones as described earlier. Although the data is available at an hourly granularity, we aggregated the discrete visits from each subzone and day into 5 time-bins spanning multiple hours: 00:00-06:00AM, 07:00-10:00AM (AM-Peak), 11:00AM-05:00PM (Off-Peak), 06:00-08:00PM (PM-Peak) and 09:00-11:00PM. Such binning allows us to investigate the utility of a larger observation window that is more attuned to the natural ``rhythm" of the city (e.g., morning peak hours, lunch hours, etc.) 

The median, first quartile, third quartile, inter-quartile range and z-score values were obtained for the weekdays and weekends, separately, for each subzone. If a particular value for a given time-bin, day and subzone was away from the first or third quartile by more than 1.5*times the IQR, it was declared an outlier. This way, a total of 864 outliers were identified across 42 weekdays, 5 time-bins and 85 subzones. Among these, it was also found that 747 were in the month of May, with more outliers observed during the first two weeks of May. The outliers were also overwhelmingly on the lower side showing diminished flow of people across many of the zones. This may  be attributed to the holiday period (1 May and 10 May being public holidays) when several of the residents could have taken a break from their regular life. A similar effect could be found around the public holiday on June 25. A key takeaway here was that although the large time windows are useful in detecting large-scale anomalies such as holidays, they fail to detect anomalies that span only over a few hours (e.g., concerts). Hence, in the following analyses, we continue to use the finest temporal granularity in order to capture both small and large events.

\subsection {Spatio-temporal Sensitivity of Individual Sources}
\label{sec:spatiotemporal}
It is our conjecture in this work that different sensor modalities, albeit all capturing aggregated mobility dynamics, will have observable differences in terms of their ability to detect urban anomalies. In this section, we seek answers for two specific questions:
\begin{enumerate}
\item{\emph{Spatial localization} - Does the event detection capability and the accuracy to which the event can be localized vary across the multiple sources?}
\item{\emph{Temporal localization} - Are the sources capable of detecting the onset of an event (closer to the start of an event when the crowds draw closer to the event venue) as it occurs?}
\end{enumerate}

\begin{figure*}
    \centering
    \begin{subfigure}[b]{0.475\textwidth}
        \centering
        \includegraphics[width=\textwidth]{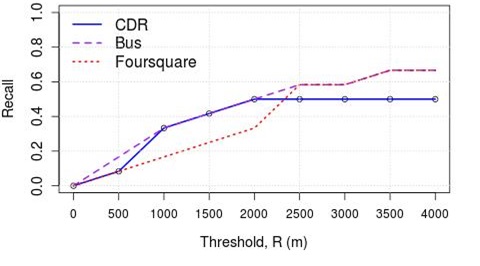}
        \caption{Outlier declared with $z-$score $\geq$ 3, at the event start hour.}   
        \label{fig:sg-h}
    \end{subfigure}
    \hfill
    \begin{subfigure}[b]{0.475\textwidth}  
        \centering 
        \includegraphics[width=\textwidth]{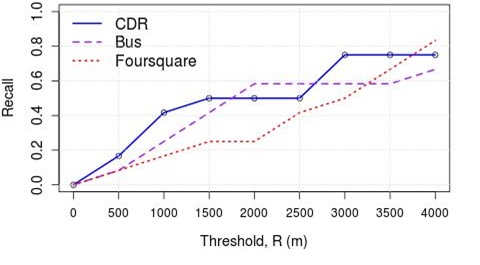}
        \caption{Outlier declared with $z-$score $\geq$ 3, at an hour prior to the event start.}    
        \label{fig:sg-hminus}
    \end{subfigure}
    \vskip\baselineskip
    \begin{subfigure}[b]{0.475\textwidth}   
        \centering 
        \includegraphics[width=\textwidth]{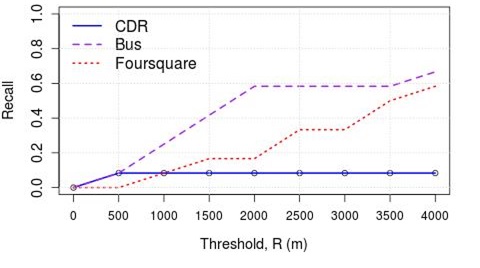}
        \caption{Outlier declared with Seasonal Hybrid ESD, at the event start hour.}    
        \label{fig:sg-h-esd}
    \end{subfigure}
    \quad
    \begin{subfigure}[b]{0.475\textwidth}   
        \centering 
        \includegraphics[width=\textwidth]{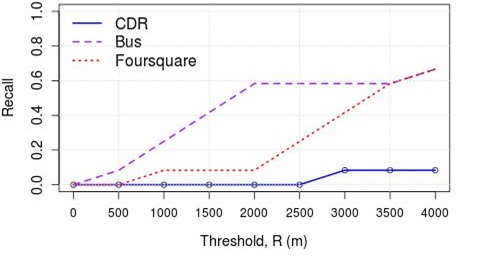}
        \caption{Outlier declared with Seasonal Hybrid ESD, at an hour prior to the event start.}    
        \label{fig:sg-hminus-esd}
    \end{subfigure}
    \caption{The trade-off between event recall and localization accuracy during (1) the event start hour and (2) and prior to the start, using the static threshold-based approach vs. ESD-based approach, on the \textbf{Singapore} datasets.} 
    \label{fig:sg-localization}
\end{figure*}

\begin{figure*}
    \centering
    \begin{subfigure}[b]{0.475\textwidth}
        \centering
        \includegraphics[width=\textwidth]{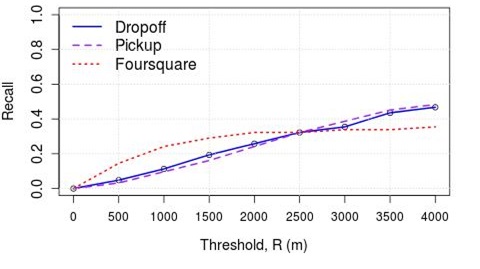}
        \caption{Outlier declared with $z-$score $\geq$ 3, at the event start hour.}   
        \label{fig:sg-h}
    \end{subfigure}
    \hfill
    \begin{subfigure}[b]{0.475\textwidth}  
        \centering 
        \includegraphics[width=\textwidth]{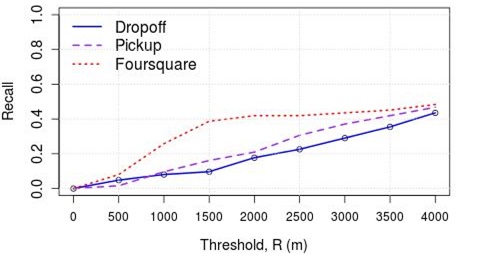}
        \caption{Outlier declared with $z-$score $\geq$ 3, at an hour prior to the event start.}    
        \label{fig:sg-hminus}
    \end{subfigure}
    
    \vskip\baselineskip
    \begin{subfigure}[b]{0.475\textwidth}   
        \centering 
        \includegraphics[width=\textwidth]{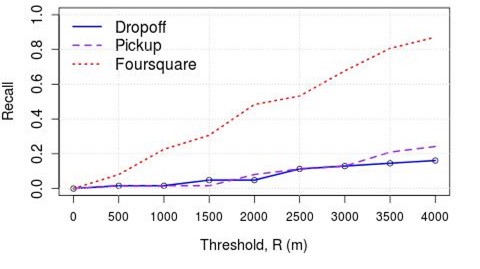}
        \caption{Outlier declared with Seasonal Hybrid ESD, at the event start hour.}    
        \label{fig:sg-h-esd}
    \end{subfigure}
    \quad
    \begin{subfigure}[b]{0.475\textwidth}   
        \centering 
        \includegraphics[width=\textwidth]{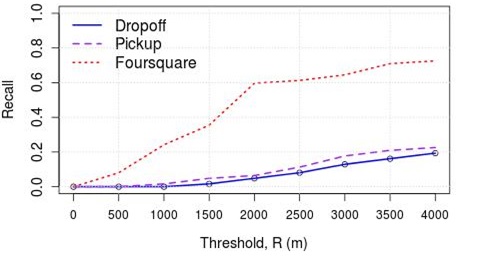}
        \caption{Outlier declared with Seasonal Hybrid ESD, at an hour prior to the event start.}    
        \label{fig:sg-hminus-esd}
    \end{subfigure}
    \caption{The trade-off between event recall and localization accuracy during (1) the event start hour and (2) and prior to the start, using the static threshold-based approach vs. ESD-based approach, on the New York City datasets.} 
    \label{fig:nyc-localization}
\end{figure*}

We measure the detection accuracy in terms of recall (i.e., proportion of events detected out of the known events (e.g., in Table~\ref{tab:events} for Singapore). We declare that an event is recalled by a source if the following criteria is met: (1) a location $s$ is an outlier for the day type $d$ and window $w$ corresponding to the event date and time, and (2) $s$ is within a radius $R$ from the event venue. To understand the temporal bias, we vary window $w$ as the same hour as the event start time, and an hour prior to that. We vary $R$ between 0 and 4000 meters. In Figure ~\ref{fig:sg-localization} and ~\ref{fig:nyc-localization}, we plot the distance threshold $R$ on $x-$axis and the recall on the $y-$axis for each source, for Singapore and NYC, respectively. We make the following observations:
\begin{enumerate}
\item{In Singapore, we note that the physical sensors, i.e., CDR and Bus Arrival, consistently perform better with the $z-score$ based anomaly declaration. Nearly 40\% of the events were detected with a localization error of less than 1.5 $km$ with both CDR and Bus loading level as observed during the starting hour of the event. }
\item{All three sensors show predictive capability with the events being detected an hour earlier than the scheduled start time -- however, we note that the recall is significantly better for the physical sensors. For instance, we observe a 10\% improvement in detection for CDR, for the same distance threshold of 1.5 $km$.}
\item{However, the detection capability drops significantly with the ESD-based approach for the physical sensors, whilst the performance of the social sensor is seen to have improved marginally. }
\item{Unlike the observations in Singapore, venue check-ins performed the best in NYC, capturing at least 30\% of the events within 1.5 $km$ of the event venue during the event start hour. Consistent with our observations from Singapore, the detection capability is higher during the hour prior to the event. This validates our earlier hypothesis that the nature of people in arriving \emph{early} for an event can in fact help detect events at their onset.}
\item{Like Singapore, we see that, even in NYC, the physical sensors weaken in their ability with the ESD-based approach whilst the social sensor sees a significant boost. We believe that the \emph{bursty} nature of social media check-ins/posts, i.e., low activity during normal days, and high bursts of activity during unusual days, makes it a better candidate for the ESD-based approach.}
\item{We further note that the detection accuracy of the social sensor is starkly less in Singapore as opposed to with the NYC data -- we believe that this may be due to the popularity of the platform during the initial years of its existence which coincides with the period of observation of the NYC setting. This may have resulted in denser check-in behavior during that time which may have influenced the observations. Understanding the impact of data sparsity, we believe, is an open question which we defer to future work for further analysis.}
\end{enumerate}

\subsection{Fusing Multiple Sources}
\begin{figure*}
    \centering
    \begin{subfigure}[b]{0.475\textwidth}
        \centering
        \includegraphics[width=\textwidth]{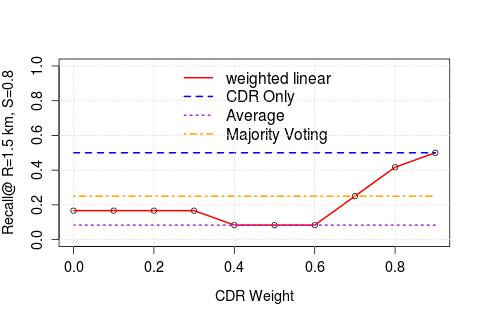}
        \caption{Singapore.}   
        \label{fig:fusion-sg}
    \end{subfigure}
    \hfill
    \begin{subfigure}[b]{0.475\textwidth}  
        \centering 
        \includegraphics[width=\textwidth]{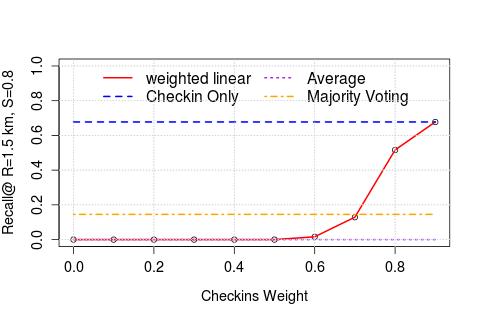}
        \caption{New York City}    
        \label{fig:fusion-nyc}
    \end{subfigure}
    \caption{Recall performance with baseline sensor fusion techniques with distance threshold, $R = 1.5 km$, and threshold on normalized $z-$ score, $s = 0.6$.} 
    \label{fig:fusion}
\end{figure*}

\begin{figure*}[b]
\centering
\includegraphics[width=7in]{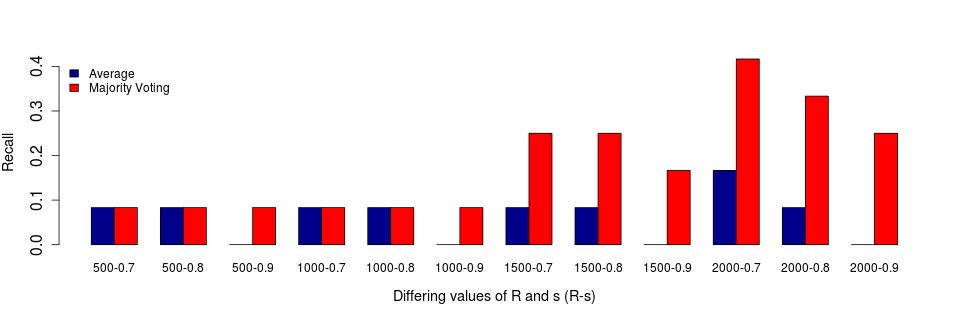}
\caption{Recall performance for varying $R$ and $s$ thresholds for Singapore.}
\label{fig:paramtuning} 
\end{figure*}

Finally, we investigate the efficacy of fusing the disparate sensor sources in the context of event detection, using baseline methods. Following from the previous analysis, we now quantify the recall performance of the individual sensors against combinations of sensors in detecting events based on the $z-$ scores. We fix the distance threshold to $R = 1.5 km$, and the threshold for declaring an observation as anomaly as $S = 0.8$ on the \emph{normalized} $z-$ scores (effectively, scaling the absolute values of the scores to between 0 and 1). In Figure~\ref{fig:fusion}, we plot the recall performance on the $y-$axis whilst the $x-$axis represents the mixture weight for the \emph{best performing sensor} (i.e., CDR for Singapore, and Check-ins for New York City) assuming a weighted linear combination. Here, we fix the weight of Check-ins and Dropoffs to 0.1, for Singapore and NYC, respectively, due to their low performance. Whilst the dashed line represents the performance using the \emph{best individual sensor} alone and the dotted purple line represents the performance of the arithmetic mean of across the three sources. Majority voting, represented by the dashed orange line, declares an anomaly when at least $k$ out of $N$ channels, individually, observe anomalies. In our settings, $k = 2$ and $N=3$, meaning at least two out of the three individual sensor modalities should have observed an anomaly. We make the following key observations:
\begin{enumerate}
\item{In both settings, it is clear that a simple fusion techniques such as an \emph{average} sensor which shows the weakest performance, does not suffice, and that this application requires more sensible forms of fusion.}
\item{By repeating the analysis over different values for $R$ (ranging between 500 m to 2 km) and $s$ (0.7-0.9), we find that fusing at the decision level (e.g., majority voting over decisions made independently per sensor) performs better than fusing at the data-level (e.g., averaging of outlier scores across the multiple sensors) in the settings we consider. We plot the recall performance observed in the case of Singapore in Figure~\ref{fig:paramtuning}.}
\item{Whilst the overall recall of majority voting is significantly less than that of the individual sensor, we believe that this also can help reduce false positives.}
\end{enumerate}


\section{Discussion}
\label{sec:discussion}
Whilst the investigations carried out in this work are rudimentary, we find that the preliminary findings to be encouraging. Event detection systems such as that described in this work can be useful to a multitude of users.  Clearly, such timely event detection is valuable in homeland security or civil defence operations, where the authorities can identify potential anomalies or disturbances proactively and activate an appropriate response protocol. In addition, such event detection has a wide variety of civilian/municipal uses as well. The general public will be able to query for current events related to their interests (e.g., condo launch) or by locality (e.g., events in a specific neighborhood). This allows for serendipitous exploration of the event landscape, as an alternative to advertised and/or ticketed events on channels such as SISTIC\cite{SISTIC}. Similarly, being able to visualize the impact of a road incident allows commuters to change their routes such that they avoid congested and possibly affected streets. On the other hand, for planning authorities, a generative model of spatiotemporal impact of events based on systematic understanding of events via multiple modalities, may help reveal previously unseen bottlenecks or unexpected anomalies that an event may cause. We next briefly describe some of the limitations in our current studies and plans for future work.

\textbf{Current Limitations and Future Work: }One of the key drawbacks in studies in event detection is the unavailability of universal ground truth of events; i.e., although effort is made to scrape/extract events from portals, news media and social media for evaluation, it is impractical to generate an exhaustive list. As a result, most events captured are inherently \emph{large} events which naturally get advertised more on multiple mediums. This also leads to problems in evaluation in that we are only able to quantify the recall of detection, and not the corresponding \emph{precision} (i.e., whether or not our system falsely detected an event). A related problem is that without complete ground-truth, it also becomes intractable to systematically tune the algorithm parameters. For instance, we have chosen the threshold on $z-$ scores as 3, or in the case of the ESD-based approach, we have imposed that the top-2\% of the observations are anomalies. However, it is not clear whether increasing or decreasing these thresholds will, in fact, achieve better quality detection. Measures need to be taken for more comprehensive ground truth collection as well as in finding alternative quantification of performance.

In addition, although presented as a component in the proposed framework, our current work does not evaluate the quality of the semantic annotation of detected events in the current study. For example, the word cloud in Figure~\ref{fig:combined} just represents the key Twitter words, but does not perform analyses (e.g., similar to Tf-Idf) to determine which words are unusual or uncommon. The semantics can be useful in multiple ways; for e.g., in disambiguating between multiple simultaneous, small scale events occurring in an area as opposed to a single, large-scale event. 

As we observe in our evaluation, a major takeaway is that traditional baseline fusion methods such as majority voting lead to poor performance. This clearly warrants the need for more carefully constructed methodologies for fusion that take in to account the spatiotemporal sensitivities of individual sensor modalities that we observe and the granularity at which data is available. One particular direction of interest is the possibility of embedding the city's natural topological network (e.g., the street network) into the analytical framework. This is an open question that we aim to tackle in future work.
\section{Related Work}
\label{sec:related}
Detecting anomalies in urban mobility patterns from physical sensors such as GPS traces and traffic cameras \cite{chawla2012inferring, liu2011discovering}, and CDR \cite{widhalm2015discovering, yin2017generative} is a well-studied topic in the context of optimizing traffic related infrastructure. CDR data have also been used to detect unusual urban events (e.g., elections, emergency events, etc. \cite{dong2015inferring, gundogdu2016countrywide}). Previous works on anomaly detection in transportation have looked at varied aspects of the problem including the detection of and finding the root causes of the anomalies. Pang et al. \cite{Pang:2011} detect anamolous regions using Likelihood Ratio Tests and, Liu et al. \cite{Liu:2011} proposed a formulation for detecting anomalous road blocks using observed minimum distortion and associating causality using frequent subtree mining. Further still, Chawla et al. \cite{Chawla:2012} use Principal Component Analysis to identify anomalous road sections. Further, in Jayarajah et al. \cite{jayarajah2015event} and Nayak et al. \cite{nayak2015exploring}, the authors show that similar techniques can be used to detect anomalies based on occupancy levels indoors (e.g., campus setting) as well as outdoors (e.g., city-scale).

Detecting events using social media is a particularly active area and has been since the initial work of Sakaki et al. \cite{Sakaki2010oYutaka} on detecting and tracking Earthquakes from user posted information on Twitter.  Due to the low fraction of geocoded posts from Twitter, recent works have studied Location-based Social Networks (e.g., Georgiev et al.\cite{GeorgievNM14b}) and image-rich platforms that are better geotagged (e.g., Jayarajah et al. \cite{jayarajah2015social, jayarajah2016can}).

However, these studies are largely unimodal in nature. A number of works that exploit multimodality have emerged; however, these focus predominantly on anomalies related to traffic (e.g., accidents). For example,  Wang et al. \cite{wang2016estimating} utilized the combination of sparsely available GPS data and Tweets to measure the congestion along road segments. Giridhar et al. \cite{giridhar2016clarisense+} make the case for identifying root causes for sensor anomalies using social media data. Recently, multimodal sensing approaches have been attempted for urban event detection. Konishi et al. \cite{konishi2016cityprophet} use a two-step modeling process to predict irregularities (e.g., large scale events) from multimodal data. However, this relies on App usage data that is only predictive of planned/anticipated events of large scale. Works such as Gao et al. \cite{Gao:2012} and Misra et al. \cite{misra2014socio} acknowledge the need for systematic fusion of multimodal data sources and discuss possible frameworks, as well as several challenges and opportunities.

\section{Conclusion} 
In this work, we investigated the ability of multimodal urban data sources in detecting events both spatial and temporally, across two Metropolitan cities: Singapore and New York City. We specifically considered 3 distinct types of urban data sources: taxi \& bus trajectories, social media feeds (from Foursquare and Twitter) and CDR data obtained from telecom companies. Our results show that different sensing modes do have unique properties and are not inherently \emph{substitutable}. Moreover, the results from our preliminary investigations are encouraging and show that appropriate fusion of the inferences from each sensing mode is likely to improve both the accuracy (recall) of event detection and permit better spatiotemporal localization of such events. However, we acknowledge the need for smarter sensor fusion techniques that account for the disparities in the spatiotemporal sensitivity of individual sensor modalities.

\acknowledgments 
This material is based on research sponsored in part by 
the U.S. Army International Technology Center Pacific (ITC-PAC), under Contract No. FA5209-17-C-0006, and partially supported by the National Research Foundation, Prime Minister's Office, Singapore under its International Research Centres in Singapore Funding Initiative. The U.S. Government is authorized to reproduce and distribute reprints for Governmental purposes notwithstanding any copyright notation thereon. 

\bibliographystyle{spiebib}
\bibliography{main} 

\end{document}